\documentstyle[preprint,tighten,aps,epsfig]{revtex}
\begin{document}
\draft

\title{Exchange currents in octet hyperon charge radii}  
\author{Georg Wagner$^{1,2}$\footnote{supported by a postdoctoral 
          fellowship of the Deutsche Forschungsgemeinschaft (DFG)
          under contract number Wa1147/1-1.  \\
          Electronic address: gwagner@physics.adelaide.edu.au},
        A.\ J.\ Buchmann$^2$\footnote{Electronic address: 
          alfons.buchmann@uni-tuebingen.de}, and 
        Amand Faessler$^2$\footnote{Electronic address: 
          amand.faessler@uni-tuebingen.de}}
\address{$^1$Centre for the Subatomic Structure of Matter (CSSM),
         University of Adelaide, Australia 5005}
\address{$^2$Institut f\"{u}r Theor.\ Physik, Universit\"{a}t T\"{u}bingen,
         Auf der Morgenstelle 14, 72076 T\"{u}bingen, Germany}
\date{\today}

\maketitle

\begin{abstract} 
  Octet hyperon charge radii are calculated in a chiral constituent quark 
  model including electromagnetic exchange currents between quarks.
  In impulse approximation one observes a decrease of the hyperon charge 
  radii with increasing strangeness. 
  This effect is reduced by exchange currents. 
  Due to exchange currents, the charge radius of the negatively charged 
  hyperons are close to the proton charge radius.
\end{abstract}

\pacs{1998 PACS number(s): 14.20.Jn, 13.40.Gp, 12.39.-x, 11.30.Rd} 
 
 \def\shiftleft#1{#1\llap{#1\hskip 0.04em}}
 \def\shiftdown#1{#1\llap{\lower.04ex\hbox{#1}}}
 \def\thick#1{\shiftdown{\shiftleft{#1}}}    
 \def\b#1{\thick{\hbox{$#1$}}}
With the advent of new facilities, charge radii of some hyperons are now under
experimental investigation \cite{Hei95}.
Different models of the nucleon \cite{Buc91,Chr96,Lu98}
 show that the negative mean square (ms) charge radius of the neutron  
$r_n^2 = -.113(3)$ fm$^2$ \cite{Kop95}  
is due to nonvalence quark degrees of freedom in the nucleon.  
On the other hand, the rms charge radius of the proton 
$r_p = 0.862(12)$ fm \cite{Sim80} is largely determined by the constituent 
quark core and the charge radius of the constituent quarks. 
The concept of two-body currents as advocated for example 
in Ref.\ \cite{Buc91} describes this situation precisely. Exchange currents 
allow to relate $r_n^2$, the quadrupole moment  of the $\Delta$-resonance 
and the $\gamma N\leftrightarrow\Delta$ transition  E2 moment 
to the $\Delta$-nucleon mass-splitting.  
The SU$_F$(3) flavor extension of this concept has been
given in Ref.\ \cite{Wag95}, and has been applied to the magnetic moments 
and the magnetic radii of all octet baryons with success. 
Recently, the effect of two-body currents has been studied 
for the radiative hyperon decays \cite{Wag98},
where they are found to be crucial for the E2/M1 ratios.

The purpose of this short article is to explore how exchange currents affect
the charge radii of the octet hyperons. 
The one-body (impulse approximation) as well as the two-body
currents are sensitive to SU$_F$(3) flavor symmetry breaking and 
strangeness suppression. 
The comparison of our model predictions and data may reveal the relevant 
two-body interactions and
currents that should be included in an effective description of low-energy QCD.
Charge radii have been studied for example in \cite{Pov90} by relating strong 
interaction and electromagnetic radii, in relativised quark models 
\cite{Bar90,War91}, the semi-bosonized SU(3) NJL model \cite{Kim96}, or
three-flavor extended Skyrme models \cite{Kun90,Gob92,Sch92}.

As a consequence of the spontaneously broken chiral symmetry of low-energy QCD,
constituent quarks emerge as the relevant quasi particle degrees of freedom
in hadron physics.
The pseudoscalar (PS) meson octet represent the corresponding Goldstone bosons.
A Hamiltonian that models this phenomenon is for three quark flavors given by
\begin{equation}
  H = \sum_{i=1}^{3} \big( m_i+ {{\bf p}^2_i\over 2m_i} \big)
  - {{\bf P}^2\over 2M} 
  + \sum_{i<j}^{3} \left( V^{\rm{Conf}}({\bf r}_i,{\bf r}_j) 
                        + V^{\rm{PS}}({\bf r}_i,{\bf r}_j)  
                        + V^{\rm{OGE}}({\bf r}_i,{\bf r}_j)  \right) \; .
\label{eq:ham}
\end{equation}
 For simplicity, we use a quadratic confinement potential $V^{\rm{Conf}}$. 
The radial form of the confinement potential is according to our experience 
not crucial for the discussion of hadronic ground state properties.
We will discuss the sensitivity of our results on different types of 
confinement interactions, e.\ g.\ linear confinement, elsewhere.  
The residual interactions comprise one-gluon exchange $V^{\rm{OGE}}$ in the 
common Fermi-Breit form without retardation corrections \cite{deR75}, and  
chiral interactions $V^{\rm{PS}}$ due to the octet of PS-mesons 
($\pi , K, \eta$) coupling to constituent 
quarks (furthermore the $\sigma$-meson is included as  chiral partner of the 
pion, while heavier scalar partners of the kaon and $\eta$ are neglected).

We use spherical $(0s)^3$ oscillator states with explicit flavor 
symmetry breaking ($m_s\neq m_d=m_u$) and SU$_{SF}$(6) spin-flavor
states for the baryon wave functions.
 For the constituent quark masses we use 
$m_u$=$m_N/3$=313 MeV and $m_u/m_s$=0.55. The effective quark-gluon coupling, 
the confinement strength and the wave 
function oscillator parameter are determined from the baryon masses.
A good description for the octet and decuplet ground state hyperons
has been obtained.
Hamiltonian (\ref{eq:ham}), as well as the wave function basis and the 
parameter fitting to the hyperon masses is described in 
Refs.\ \cite{Wag95,Wag98}.

\begin{figure}[h]
  {\epsffile{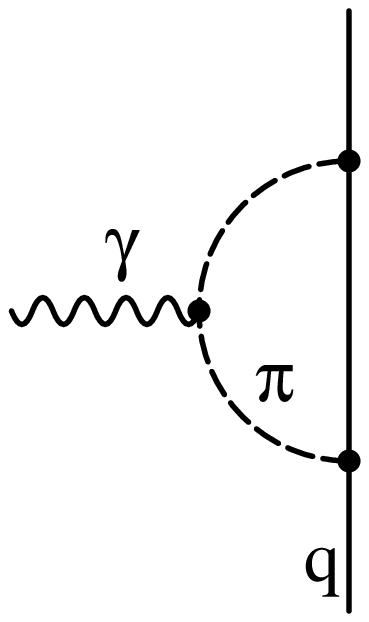}}
\caption{Pion loop contribution to the electromagnetic form factors of the
         constituent quark. Vector meson dominance relates the finite
         electromagnetic size of the constituent quark to the vector 
         meson mass $r_{\gamma q}^2 \simeq 6/m_\rho^2$.}
\label{figure:quark}
\end{figure}

Constituent quarks are obtained through the dressing of current quarks by
their strong interactions.
 For an electromagnetic probe they appear as extended objects (see for example
Refs.\ \cite{Vog90} for further discussions). 
The quark currents are screened by quark-antiquark polarization effects.
This cloud of the constituent quark is dominated by $q\bar q$ pairs with 
pseudoscalar meson quantum numbers as sketched in Fig.\ \ref{figure:quark}, 
leading to the vector meson dominance picture.
In our model the electromagnetic size of the constituent quark 
is described by a monopole form factor \cite{Buc91}
\begin{equation}
  F_{\gamma q}({\bf{q}^2}) = \frac{1}{1 + {\bf{q}^2}r_{\gamma q}^2/6} \; .
\label{eq:cqs} 
\end{equation}
Guided by vector meson dominance, we choose an electromagnetic constituent 
quark size of $r_{\gamma q}^2 = (0.6 {\rm{fm}})^2 \simeq 6/m_\rho^2$, 
independent of quark flavor for simplicity. 

\begin{figure}[h]
  {\epsffile{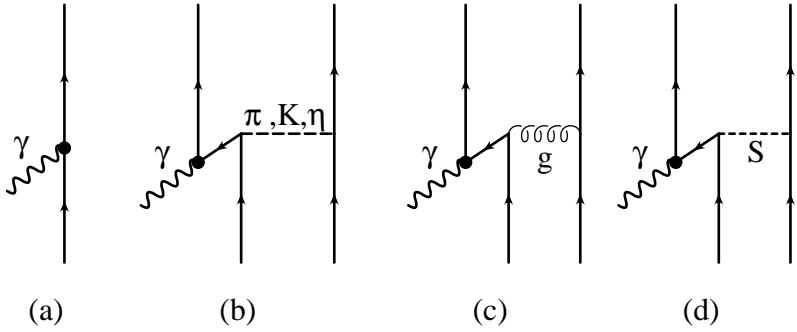}}
\caption{(a) Impulse approximation, (b) PS-meson pair current 
         ($\pi , K, \eta$), (c)  gluon-pair current, and (d) 
         scalar exchange currents (confinement and $\sigma$-exchange).
         The black dot at the photon-quark vertex refers to the constituent
         quark dressing.}
\label{figure:currents}
\end{figure}

Charge radii are calculated from the charge form factor
\begin{equation}
  F_{\rm{C0}}^B({\bf{q}^2}) = \sqrt{4\pi} \;
  \langle B \vert \frac{1}{4\pi} \int d\Omega_{\bf{q}} \cdot
  Y_{\rm{00}}(\hat{\bf{q}}) \,\rho ({\bf{q}}) \,\vert B \rangle \; ,
\label{eq:ff} 
\end{equation}
where the electromagnetic one-body and two-body charge densities 
$\rho ({\bf{q}}) = \sum_i \rho_i^{\rm{Imp}} ({\bf{q}}) + 
                 \sum_{i<j} \rho_{i,j}^{\rm{Exc}} ({\bf{q}})$
corresponding to Hamiltonian (\ref{eq:ham}) are constructed
by a non-relativistic reduction of the Feynman diagrams 
\cite{Buc91,Wag98} shown in Fig.\ \ref{figure:currents}. 
 Form factor (\ref{eq:cqs}) multiplies the charge form factor (\ref{eq:ff}).

Charge radii are defined as the slope at ${\bf q}^2$=0.
\begin{equation}
  r_B^2 = -6 \left. 
  {\frac{\partial}{\partial {\bf{q}}^2}F_{\rm{C0}}^B({\bf{q}}^2) } 
  \right\vert_{{\bf{q}}^2=0}
  \equiv r_{\rm{Imp}}^2 + r_{\rm{Exc}}^2    
\label{eq:rad} 
\end{equation}
(and are normalized to the hyperon charge where necessary).
The various one- ($r_{\rm{Imp}}^2$) and two-body ($r_{\rm{Exc}}^2$) 
contributions to the charge radii are given in Table  \ref{table:radii}.

The neutron charge radius is well described by gluon and pseudoscalar 
meson exchange currents. 
This result is independent of the constituent quark size and the Hamiltonian 
parameters, but is strongly sensitive to the harmonic oscillator parameter
\cite{Buc91}, fixing it to $b  = 0.61 \, {\rm{fm}}$ used in our calculation.
For the proton charge radius, we find $r_p = 0.807 \, {\rm{fm}}$ with 
$r_{\gamma q}^2 = 0.36 \, {\rm{fm}}^2$.
Leaving other parameters fixed, we obtain for 
$r_{\gamma q}^2 = 0.42 \,{\rm{fm}}^2$ a charge radius of
$r_p = 0.848 {\rm{fm}}$ (giving good proton and neutron magnetic radii),
and for $r_{\gamma q}^2 = 0.49 \,{\rm{fm}}^2$ we obtain 
$r_p = 0.889 \,{\rm{fm}}$.
Recently, the experimental proton charge radius is under scrutiny.
While an atomic physics experiment suggests a value of 0.890(14) fm 
\cite{Ude97}, the experimental value from elastic $ep$ scattering 
is 0.862(12) fm \cite{Sim80}. 
A dispersion relation analysis \cite{Mer96} gives 
a smaller value of $r_p$=0.847(9) fm. 

Our results for the proton and neutron charge radii, the agreement with other 
models \cite{Chr96,Lu98} concerning the dynamical origin of the negative 
neutron charge radius, and our results for the octet baryon magnetic moments 
\cite{Wag95} give us confidence to extend the calculation of exchange current 
effects in charge radii to the entire baryon octet.
In the following, we discuss our results for all octet hyperon charge radii 
relative to the proton and neutron charge radii.

In impulse approximation, SU$_F$(3) symmetry breaking, i.\ e.\ the fact that
$m_u/m_s=$0.55--0.6 as suggested by octet magnetic moments \cite{Wag95}, 
leads to a reduction of the charge radii with increasing strangeness 
content of the hyperon, and as a consequence 
the proton, $\Sigma^-$, and $\Xi^-$ have successively smaller charge radii. 
This strangeness suppression almost disappears when exchange currents are 
included.  
As can be seen from Table \ref{table:radii}, the effect is largely due to the 
SU$_F$(3) symmetry breaking in the pseudoscalar meson exchange currents.
A relatively strong reduction of the proton charge radius by the pion cloud 
occurs, whereas the corresponding decrease in the $\Sigma^-$ (or $\Xi^-$) is 
mainly due to the kaon cloud which is a factor of three smaller in magnitude
due to the larger kaon mass. 
Overall, the pseudoscalar mesons reduce the charge radii of $p,\Sigma^-, \Xi^-$
by -.057 fm$^2$, -.029 fm$^2$, and -.022 fm$^2$ respectively.
Due to the pseudoscalar meson cloud in the hyperons, the charge radius of the 
$\Sigma^-$ (or $\Xi^-$) turns out to be as large as the proton charge radius. 
The gluon- and scalar exchange current contributions, which 
when considered separately show strong strangeness suppression, are in 
their sum with -.027 fm$^2$, 
-.020 fm$^2$, and -.024 fm$^2$ for $p,\Sigma^-, \Xi^-$ almost constant. 
It is important to note that for the hyperons the contributions of the kaon or
$\eta$ are of the same magnitude as the pion exchange currents.
The spin-flavor matrix element of the pion pair charge operator
for the $\Sigma^-$ is 9 times smaller than for the proton.  

In impulse approximation as well as after inclusion of exchange currents the
$\Sigma^+$ charge radius is significantly larger than the proton 
ms charge radius.  
This is in line with the intuitive picture that the probability of finding the 
heavier strange quark is larger at the center of the hyperon whereas the 
probability distribution of the lighter u-quarks extends to larger distances. 
On the other hand, the CQM overpredicts the $\Sigma^+$ magnetic moment 
\cite{Wag95}, and doubts about the $\Sigma^+$ wave function in quark models 
\cite{Lip81} make the situation unclear.
In the SU$_F$(3) limit, the wave functions and radii of the $\Sigma^+$ and 
the proton would be identical.

In most cases exchange current contributions reduce the charge radii as 
compared to the impulse approximation. 
At first sight this seems to be counter-intuitive,
because one expects the PS meson cloud to increase the charge radius.
However, closer inspection shows that the reduction is physically plausible.
First, in impulse approximation the proton charge form factor drops off too 
fast compared to the dipole fit at moderate momentum transfers 
of $q^2\sim {\rm{3 fm}}^{-2}$.
After inclusion of the lowest order relativistic corrections in the charge 
density operator (due to exchange currents) the charge 
form factor compares more favorably with the empirical 
dipole fit even up to larger momentum 
transfers ($q^2\sim {\rm{10 fm}}^{-2}$). 
The slower decrease of the form factor when exchange currents are included 
is manifest at $q^2=0$ in the smaller charge radius of the proton. 
The same result is obtained for almost all octet hyperon radii.
Second, one has to keep in mind, that a part of the
cloud effects are already included in impulse approximation, where our modeling
of the constituent quark dressing through Eq.\ (\ref{eq:cqs}) contributes 
$r_{\gamma q}^2$=0.36 fm$^2$ for charged hyperons.
 
Third, the spin S=0 quark pairs in the hyperons lead to smaller 
charge distributions.
As we mentioned before, the negative neutron charge radius is due to
the spin-dependent exchange currents between quarks, giving negative
contributions for all S=0 quark pairs in the nucleon. 
The same effect holds true for all octet hyperons.
For all decuplet hyperons, in which two quark spins are coupled to S=1, 
the net effect of exchange currents is to increase the charge 
radii obtained in impulse approximation 
(except for the $\Delta^0$, for which the charge form factor is 0 even 
after inclusion of exchange currents, due to isospin symmetry).
Decuplet-octet splittings in the electromagnetic radii occur through the spin
carrying exchange current degrees of freedom in the hyperons. 
This is closely related to the decuplet-octet mass splittings caused
by the spin-dependent potentials.

In our model, the neutron is the only baryon with a negative ms charge radius,
while $\Lambda$ and $\Xi^0$ have a negative square charge radius
only in the SU$_F$(3) limit. 
In that limit impulse contributions vanish, and the charge radius of 
$\Xi^0$ would be identical to the neutron charge radius. 
After symmetry breaking, the impulse approximation radii for 
$\Sigma^0$, $\Lambda$, and $\Xi^0$ are nonzero and positive. 
The exchange current contributions to the $\Lambda$ and $\Xi^0$ charge radii 
are negative, and drastically reduce the 
SU$_F$(3) symmetry breaking of the impulse approximation. 
The confinement exchange current contributions to the neutral hyperon radii are
rather large despite the fact that they are only due to SU$_F$(3) symmetry 
breaking.

In comparison with other works, the orderings of the charge radii in Ref.\ 
\cite{Kim96} agree with our result, as can be seen from Table \ref{table:comp}.
However, they suggest a negative ms charge radius of $\Lambda$ and $\Xi^0$. 
The $\Xi^0$ radius is positive in most models despite being negative 
(the same as the neutron) in the SU$_F$(3) limit.
The strongest flavor symmetry breaking is observed in \cite{Sch92}, while
Ref.\ \cite{Kim96} exhibits only a small effect.
We confirm the observation of \cite{War91} that the neutron and $\Sigma^0$ 
charge radii are similar in their absolute values, as opposed to 
most models in Table \ref{table:comp}.

The cloud effects in our model are large for the $\Lambda$.
The $\Lambda$ is interesting since various models obtain different signs for 
its charge radius.
We have seen, that for the $\Lambda$, exchange currents cancel the impulse 
approximation to a large extent. 
The $\Lambda$ charge radius appears therefore as a sensitive probe for model 
treatments of meson cloud effects. 
Some models observe (almost) identical radii for $\Sigma^0$ and $\Lambda$.
In contrast to this, the big difference between $\Sigma^0$ and $\Lambda$ 
in our model is due to the spin dependent two-body charge density operators, 
and is therefore of the same origin as the negative charge radius of the 
neutron.
                              
Our model, as well as two rather different model treatments of SU$_F$(3)
symmetry breaking \cite{Kim96,Sch92}, suggest a small strangeness suppression 
of the $\Xi^-$ charge radii of less than 10$\%$ as compared to the proton.
This is however not true for other Skyrme model results \cite{Kun90,Gob92}.
Calculations that neglect meson cloud effects \cite{Bar90} find rather large
strangeness suppression.

Summarizing, we have studied octet hyperon charge radii within a chiral 
constituent quark model including two-body exchange currents.
 For charged hyperons, exchange currents decrease the 
impulse approximation radii by less than 10$\%$.
This is consistent with our results for octet magnetic moments \cite{Wag95},
where cancellations among exchange currents give rise to 10$\%$ modifications.
SU$_F$(3) symmetry breaking effects as observed in impulse approximation  
are found to be
weakened by exchange currents both for charged and neutral octet hyperons.
The negatively charged hyperons have radii comparable to the proton charge
radius due to the pseudoscalar meson clouds surrounding the quark core.
After SU$_F$(3) symmetry breaking the neutron is the only baryon with  
negative mean square charge radius, while the $\Lambda$ and $\Xi^0$ 
radii are small and positive.
Different model predictions of hyperon charge radii vary considerably in their
treatment of SU$_F$(3) symmetry breaking, in particular with respect to 
the cloud effects. 
The forthcoming experimental results for the hyperon
charge radii provide a test of the concepts discussed in this paper, and 
will discriminate between different models. 


\begin{table}[h]
\begin{tabular}{ l | c  c  c  c  c  c  c | c  c }
       &   \rule[-2mm]{0mm}{8.mm}$r_{\rm{imp}}^2$ & $r_{\rm{g}}^2$ & 
           $r_{\rm{conf}}^2$ & $r_\sigma^2$ & $r_{\pi}^2$ & $r_{\rm{K}}^2$ & 
           $r_{\eta}^2$ & $r_{\rm{tot}}^2$ & $\sqrt{\vert r_{\rm{tot}}^2\vert }$
           \\[0.25cm] \hline    \rule[-2mm]{0mm}{8.mm}
$p$ &        .736 & .123 &-.196 & .046 & -.057 &   0   &  0   & .651 & .807 \\
$\Sigma^-$ & .691 & .054 &-.107 & .033 & -.006 & -.017 &-.005 & .643 & .802 \\
$\Xi^-$    & .633 & .001 &-.047 & .022 &   0   & -.017 &-.005 & .587 & .766  
           \\[0.25cm] \hline   \rule[-2mm]{0mm}{8.mm}
$\Sigma^+$ & .861 & .118 &-.170 & .045 & -.013 & -.009 &-.007 & .825 & .909 
           \\[0.25cm] \hline    \rule[-2mm]{0mm}{8.mm}
$n$ &         0   &-.082 &   0  &   0  & -.038 &   0   & .003 &-.117 & .342 \\
$\Lambda\leftrightarrow\Sigma^0$ & 
              0   & .057 &   0  &   0  &   0   &  .005 & .004 & .066 & .257 
           \\[0.25cm] \hline    \rule[-2mm]{0mm}{8.mm}
$\Sigma^0$ & .085 & .032 &-.034 & .006 & -.003 &  .004 &-.001 & .089 & .296 \\
$\Lambda$  & .085 &-.014 &-.034 & .006 & -.028 &  .003 & .001 & .019 & .131 \\
$\Xi^0$    & .169 &-.034 &-.039 & .008 &   0   & -.009 &-.005 & .091 & .302 
\end{tabular} 
\caption[octet charge radii]{Mean square charge radii of octet baryons. 
         The first column contains the impulse approximation result 
         $r_{\rm{imp}}^2$.
         In the next six columns the individual exchange current
         contributions are listed separately: gluon- $r_{\rm{g}}^2$, 
         confinement- $r_{\rm{conf}}^2$, and one-sigma exchange currents 
         $r_\sigma^2$, and the PS meson exchange current contributions
         $r_{\pi}^2,r_{\rm{K}}^2,r_{\eta}^2$. 
         The total hyperon charge radii $r_{\rm{tot}}^2$ are given in the 
         second to last column.
         All quantities are given in [fm$^2$], except for the last column 
         which is given in [fm]. 
         Until now, only the charge radii of proton and 
         neutron $r_p=(0.862\pm .012)\,\rm{fm}$ \cite{Sim80} and 
         $r_n^2=-(0.113\pm .003)\,\rm{fm}^2$ \cite{Kop95} have been accurately
         measured.}
\label{table:radii}
\end{table}

\begin{table}[h]
\begin{tabular}{ l | c  c | c  c  c  c  c  c }
  \rule[-2mm]{0mm}{8.mm}[fm$^2$] & $r_{\rm{Imp}}^2$ & $r_{\rm{Tot}}^2$ 
  & RQM \cite{Bar90}      & RCQM \cite{War91}         & NJL \cite{Kim96}   
  & Skyrme \cite{Kun90}   & Skyrme \cite{Gob92}     & Skyrme \cite{Sch92} 
\\[0.25cm] \hline \rule[-2mm]{0mm}{8.mm}
  $p$        & .736 & .651 & .664 &      &  .78 & .775 &  .75 &  .52  \\
  $\Sigma^-$ & .691 & .643 & .570 &      &  .75 & .751 &  .88 &  .52   \\
  $\Xi^-$    & .633 & .587 & .475 &      &  .72 & .261 &  .48 &  .47  
\\[0.25cm] \hline \rule[-2mm]{0mm}{8.mm}
  $\Sigma^+$ & .861 & .825 & .753 &      &  .79 & .964 & 1.08 &  .442 
\\[0.25cm] \hline \rule[-2mm]{0mm}{8.mm}
  $n$        &  0   &-.117 & 0    &-.081 & -.09 &-.308 & -.28 & -.057 \\
  $\Lambda\leftrightarrow\Sigma^0$ & 
                0   & .066 & --   &      &  --  & --   & --   &  --   
\\[0.25cm] \hline \rule[-2mm]{0mm}{8.mm}
  $\Sigma^0$ & .085 & .089 & .091 & .083 &  .02 & .107 & .10  & -.036 \\
  $\Lambda$  & .085 & .019 & .091 & .120 & -.04 & .107 & .10  & -.031 \\
  $\Xi^0$    & .169 & .091 & .190 &      & -.06 & .221 & .41  &  .00 
\end{tabular} 
\caption[comparison with other authors]{Comparison of our results
        with other works. The relativistic quark model of \cite{Bar90} uses a
        logarithmic confinement. Relativised constituent 
        quark model results from \cite{War91} and
        results from the semi-bosonized SU(3) NJL model \cite{Kim96} are shown
        together with Skyrme model calculations, which use the bound state 
        \cite{Kun90,Gob92} and the slow-rotor approach \cite{Sch92},
        respectively. All quantities are in fm$^2$.}
\label{table:comp}
\end{table}

\end{document}